\magnification\magstep1
\scrollmode
\overfullrule=0pt
\hfuzz=1pt

\font\tenbbb=msbm10 \font\sevenbbb=msbm7         
\font\tenams=msam10 \font\sevenams=msam7 \font\tenbm=cmmib10
\font\sevenbm=cmmib10 at 7pt \font\eightit=cmti8 
\font\eightrm=cmr8 \font\eighti=cmmi8            
\font\eightsy=cmsy8 \font\eightams=msam8

\newfam\bbbfam \newfam\amsfam \newfam\bmfam      
\textfont\bbbfam=\tenbbb \scriptfont\bbbfam=\sevenbbb
\textfont\amsfam=\tenams
\scriptfont\amsfam=\sevenams  
\textfont\bmfam=\tenbm \scriptfont\bmfam=\sevenbm

\def\k{\kappa}                      
\def\a{\alpha}                      
\def\r{\rho}                        
\def\s{\sigma}                      
\def\ji{\chi}                       
\def\vep{\varepsilon}               
\def\b{\beta}                       
\def\cite#1{{\rm[#1]}}               
\def\g{\gamma}                      
\def\d{\delta}                      
\def\Q{{\sl Q}}
\def\S{{\cal S}}
\def\G{\Gamma}
\def\x{\xi}
\def\part{\partial}
\def\P{{\cal P}}
\def\bphi{\overline{\phi}}
\def\bP{\overline{P}}


\newif\ifstartsec                   

\outer\def\section#1{\vskip 0pt plus .15\vsize \penalty -250
\vskip 0pt plus -.15\vsize \bigskip \startsectrue
\message{#1}\centerline{\bf#1}\nobreak\noindent}

\def\subsection#1{\ifstartsec\medskip\else\bigskip\fi \startsecfalse
\noindent{\it#1}\penalty100\medskip}

\def\refno#1. #2\par{\smallskip\item{\rm\lbrack#1\rbrack}#2\par}

\def\eightpoint{\normalbaselineskip=10pt 
\def\rm{\eightrm\fam0} \let\it\eightit
\textfont0=\eightrm \scriptfont0=\sevenrm 
\textfont1=\eighti \scriptfont1=\seveni
\textfont2=\eightsy \scriptfont2=\sevensy \textfont\amsfam=\eightams
\normalbaselines \eightrm
\parindent=1em}


\def\Wittsm{1}
\def\Physrep{2}
\def\Horne{3}
\def\Birmingham{4}
\def\Witcoh{5}
\def\Topmat{6}
\def\Lambert{7}
\def\Background{8}
\def\Cordes{9}
\def\Bos{10}


\rightline{FT/UCM--7--96}

\vskip 1cm

\centerline{\bf On the ``gauge'' dependence of the toplogical sigma model
                beta functions}

\bigskip

\centerline{\rm Luis Alvarez-C\'onsul\dag}
             \medskip
\centerline{\eightit Departamento de F{\'\i}sica Te\'orica, Universidad 
                     Aut\'onoma de Madrid, Cantoblanco,28049 Madrid, Spain}
\vfootnote\dag{email: {\tt Lalvarez@delta.ft.uam.es}}
\bigskip
\centerline{\rm C. P. Mart{\'\i}n*}
\medskip
\centerline{\eightit Departamento de F{\'\i}sica Te\'orica I,
                     Universidad Complutense, 28040 Madrid, Spain}
\vfootnote*{email: {\tt carmelo@eucmos.sim.ucm.es}}

\bigskip\bigskip

\begingroup\narrower\narrower
\eightpoint
We compute the dependence on the classical action ``gauge'' parameters 
of the beta functions of the standard topological sigma model in flat space. 
We thus  show that their value is a ``gauge'' artifact indeed. 
We also show that previously computed values of these beta functions 
can be continuously connected to one another by smoothly varying
those ``gauge'' parameters.
\par
\endgroup 

\bigskip
\centerline{PCAS numbers 11.10.Gh, 11.10.Kk, 11.10.Lm}
\bigskip

The topological sigma model introduced in ref. \cite{\Wittsm} is a 
particular instance of Topological Field Theory 
(see \cite{\Physrep} for a review), and it lacks, therefore, 
physical propagating local degrees of freedom. As a result, the 
observables of the model are expected to be ultraviolet finite. And yet,
the beta functions of the model for the classical action of 
ref. \cite{\Wittsm} turns out not to vanish \cite{\Horne}. 
The solution to this riddle has been advanced by the authors of 
ref. \cite{\Birmingham}. These authors suggest that a non-vanishing 
beta function is  merely a ``gauge'' artifact, which has therefore
no bearing on the value of the observables of the model. They back their 
argument by introducing a classical action for the topological 
sigma model that continuously connects, by means of a ``gauge'' parameter, say,
$\k_1$, the action of ref. \cite{\Wittsm}, which demands $\k_1\,=\,1$,
with the ``delta gauge'' action, which corresponds to $\k_1\,=\,0$. 
They then go on and compute the one-loop contributions to 
the effective action for  the ``delta gauge'' classical action. 
These contributions are ultraviolet finite so that the 
one-loop beta functions for the delta-gauge action vanish. The issue,  
however, has not been settled yet since it has not been shown that 
the non-vanishing beta functions obtained in ref. \cite{\Horne} can be 
continuously make to vanish by sending $\k_1$ to zero.
It may well happen that the theories obtained for $\k_1\,=\,1$ 
and $\k_1\,=\,0$ are not the same quantum theory in spite  
of the fact that their classical actions differ by a BRSTlike-exact term:
anomalies may turn up upon quantization. 
It is thus necessary to compute the beta functions for arbitrary values of 
$\k_1$ and show that these functions connect continuously the non-vanishing 
beta functions obtained in ref. \cite{\Horne} with the 
vanishing beta functions of ref. \cite{\Birmingham}. The purpose of 
this paper is to carry out this computation and show that the 
beta functions depend on $\k_1$ as expected.

We would like to do our computations by using the superfield formalism 
introduced in ref. \cite{\Horne}. The first issue to tackle will thus be
the existence of a superfield action that matches the action in 
ref. \cite{\Birmingham}. The latter action is obtained by setting $\k_2\,=\,1$ 
in the following equations
$$
\S(\k_1,\k_2)\,=\,- i\{\Q,{\rm V}(\k_1, \k_2)\}\; ,
\eqno (1)
$$
$$
{\rm V}(\k_1,\k_2)\,=\, \int\! d^{2}\s\;
 \r^{\a i}\Bigl( -{\k_1 \over 4} H_{\a}^{\; i}+
\k_2\, \partial_{\a} u^{i}\Bigr)\, g_{i j}\; .  
\eqno (2)
$$
Let us display the field content of the model. 
First, we introduce the fields $u^i(\s)$, 
which have conformal spin zero. $u^{i}(\s)$ describing (locally) a 
map $f$ from a Riemann surface, $\Sigma$, to an almost complex 
riemannian manifold $M$; the  almost complex extructure on $M$ being
denoted by $J^i_{\; j}$. Notice that the symbol $g_{i j}$ stands 
for the hermitian metric on $M$ with regard to $J^i_{\; j}$; 
 $\s$ denotes a point in $\Sigma$. Secondly, we define
the  anticonmuting field $\ji^{i}(\s)$ of conformal spin cero 
to be geometrically interpreted as a section of the pullback by 
$f$ of the tangent bundle to $M$; this pullback being denoted by $f^*(T)$.
We need two more fields. Let us call them  $\r^{\a i}$ and
$H^{\a i}$, respectively. They both have conformal spin one, and, they both
give rise to sections of the bundle of one forms over $\Sigma$ with values
on $f^{*}(T)$. The fields $\r^{\a i}$ and $H^{\a i}$ are anticommuting
and commuting objects, respectively, and they obey the selfduality constraints
$$
\r^{\a i}\,=\,\vep^{\a}_{\; \b} J^{i}_{\; j}\, \r^{\b j}\quad
H^{\a i}\,=\,\vep^{\a}_{\; \b} J^{i}_{\; j} H^{\b j}\; .
\eqno(3)
$$
Here $\vep^{\a}_{\; \b}$ is the complex structure of $\Sigma$, verifying
$\vep^{\a}_{\;\b} \vep^{\b}_{\;\g} = -\d^{\a}_{\;\g}$. The greek indices are
tangent indexes to $\Sigma$, they take on two values, say, $1$ and $2$.
These indexes are raised and lowered by using a metric, $h_{\a \b}$, compatible
with the complex structure $\vep^{\a}_{\; \b}$. It should be mentioned that
roman indices run from $1$ to ${\rm dim}\, M$ and that they are associated to
a given basis of $f^*(T)$.

The symbol $\Q$ denotes the BRST-like charge characteristic of cohomological
field theories \cite{\Witcoh}, which can be obtained by 
``twisting'' \cite{\Topmat} the appropriate N=2 supersymmetric field 
theory \cite{\Topmat, \Lambert}. The action of $\Q$ on the fields 
introduced above reads \cite{\Wittsm}
$$
\eqalignno{&\{\Q, u^{i}\}\,=\,-\ji^{i},\qquad \{\Q, \ji^{i}\}\,=\, 0\cr
&\{\Q, \r^{\a i}\}\,=\, i\Bigl( H^{\a i}+{1\over 2}i \vep^{\a}_{\;\b}
D_{k} J^{i}_{\; j}
\ji^{k}\r^{\b j}-i \G^{i}_{j k}\ji^{j}\r^{\a k}\Bigr)\cr
&\{\Q, H^{\a i}\}\,=\, i\Bigl( -{1\over 4}(R^{i}_{ j k l}- R^{m}_{n k l}
J^{i}_{\; m} J^{n}_{\;\; j}) \ji^{k}\ji^{l} \r^{\a j} +
{i\over 2}  \vep^{\a}_{\;\b} D_{k} J^{i}_{\; j} \ji^{k} H^{\b j}+\cr
&{1\over 4} D_{k} J^{i}_{\; m} D_{l} J^{m}_{\quad j} \ji^{k}\ji^{l}
\r^{\a j} -i \G^{i}_{j k} \ji^{j} H^{\a k} \Bigr)
\; .&(4)\cr
}
$$
In the preceding equations $\G^{i}_{j k}$ stands for the Levi-Civita
connection on $M$ and $R^{i}_{j k l}$ denotes the Riemann tensor for
this connection.

Besides the BRST-like symmetry $\Q$, the model whose 
action is displayed in eq.
(2) has at the classical level a $U(1)$ symmetry which obeys $[U,\Q]=0$.
The fields $u$, $\ji$, $\r$ and $H$ have, respectively, the following 
$U(1)$ quantum numbers: $0$, $1$, $-1$ and $0$. The action $\cal S$ 
is conformal invariant and it has $U\, =\, 0$.

Following ref. \cite{\Horne} we next introduce an anticommuting variable
$\theta$ with conformal spin zero and $U(1)$ charge charge -1, and, define
the following superfields
$$
\phi^{i}(\s, \theta)\,=\, u^{i}(\s)+i\,\theta\ji^{i}(\s)\; ,
\eqno (5)
$$
$$
P^{\a i}(\s,\theta)  = \r^{\a i} + \theta 
\,\bigl( H^{\a i}-{1\over2} i D_{k} J^{i}_{\; j}
J^{j}_{\; l}\ji^{k}\r^{\a l} - i \G^{i}_{j k} \ji^{j} \r^{\a k})\; .
\eqno (6)
$$
The superfields $\phi^i$ have boths $U(1)$ charge and conformal spin $0$.
The anticommuting superfield 
$P^{\a i}$, which is constrained by a selfduality equation 
analogous to eq. (3),
has $U(1)$ charge $-1$ and conformal spin $1$. The action of $\Q$ on the
the superfields in eqs. (5) and (6) is given by ${\part \over \part\theta}$.
 
We are now ready to establish a superspace formulation of our action.
It is not difficult to show that the  action in eq. (1) can be recast into
the following form
$$
\S(\k_1, \k_2)\,=\,\int\! d^2\s d\theta\;\Bigl( -{\k_1 \over 4}\, 
P^{\a i}D_{\theta} P_{\a}^{\; j}\,
g_{i j}(\phi) + \k_2\,P^{\a i}\part_{\a}\phi^{j}\, g_{i j}(\phi)\Bigr)\, ,
\eqno (7)
$$
where 
$D_{\theta}P^{\a i}\,=\, \part_{\theta}P^{\a i} + \part_{\theta}\phi^{j}
\G^{i}_{j k} P^{\a k}$. One recovers the superspace action of ref. 
\cite{\Horne} by setting $\k_1\,=\,\k_2\,=\,1$ in eq. (7),  
whereas $\k_1=0,\k_2=1$ corresponds to a superspace formulation 
of the ``delta gauge'' action introduced in 
ref. \cite{\Birmingham}. Notice that the action in eq. (7) has $U\,=\,0$ and
that it is superconformal invariant, as required.

Our next move will be the computation of the beta functions of our model for
generic values of $\k_1$ and $\k_2$. We shall quantize the model by
using the background field method \cite{\Background}. It is a lengthy, 
though straightforward, computation to carry out the expansion of the
action $\S(\k_1, \k_2)$ around the isolated background field configurations
$\bphi^{i}$ and $\bP^{\a i}$; the latter corresponding to the full quantum
superfields $\phi^{i}$ and $P^{\a i}$, respectively. However, to unveil the 
one-loop ultraviolet divergent structure of our model, and carry out its
renormalization, we need only to consider the following contributions
$$
\overline{\S}(\k_1, \k_2)\,=\,\int\! d^2\s d\theta\;\Bigl( -{\k_1 \over 4}\, 
\bP^{\a i}D_{\theta} \bP_{\a}^{\;\, j}\,
g_{i j}(\bphi) + \k_2\,\bP^{\a i}\part_{\a}\bphi^{j}\, g_{i j}(\bphi)
\Bigr)\, ,
\eqno (8)
$$
$$
\S_{\rm prop}(\k_1, \k_2)\,=\,\int\! d^2\s d\theta\;\Bigl( -{\k_1 \over 4}\, 
\P^{\a i}D_{\theta} \P_{\a}^{j}\,
g_{i j}(\bphi) + \k_2\,\P^{\a i}D_{\a}\x^{j}\,  g_{i j}(\bphi)
\Bigr)\, ,
\eqno (9)
$$
$$
\eqalignno{
&\S_{\rm int}(\k_1, \k_2)\,=\, \int\! d^2\s d\theta\,\Bigl\{\Bigl[
 -{\k_1 \over 4}\, 
\bP^{\a i}D_{\theta} \bP_{\a}^{j}\,\Bigl(
-{1\over 2} J^{m}_{\quad\!  i} D_k D_l J_{j m} + {1\over 3} R_{iklj}\cr
& - {1\over 2}
D_k J^m_{\quad\! i} D_l J_{j m} \Bigr)+
\k_2\,\bP^{\a i}\part_{\a}\bphi^{j}\,\Bigl( 
-{1\over 4} J^m_{\quad\! i} D_k D_l J_{jm} + {7\over 12} R_{iklj} \cr 
&+ {1\over 12}
R_{mkln}J^n_{\quad\! i} J^m_{\quad j} 
- {1\over 8} D_k J^m_{\quad\! i} D_l J_{jm}\Bigr)\Bigr]\x^k \x^l+\cr
&J^{m}_{\quad\! k}D_l J^i_{\; m}\,\Bigl(
{\k_1 \over 2}\,\bP^{\a k}D_{\theta} 
\P^{\; j}_{\a} - {\k_2 \over 2}\,\P^{\a k}\part_{\a}\bphi^{j}\Bigr)
\, g_{i j}(\bphi) \x^l+\k_2\,
D_k \bP^{\a i} D_{\a}\x^j\, \x^k\,g_{i j}(\bphi)\Bigr\}
\; . & (10)\cr 
}
$$
The fields $\x^i$ and $\P^{\a i}$ embody the quantum fluctuations around the
background fields $\bphi^i$ and $\bP^{\a i}$, respectively. In the 
background field quantization procedure, one integrates over
$\x^i$ and $\P^{\a i}$  to obtain the effective action \cite{\Horne}.

Integrating out the fields $\x^i$ and $\P^{\a i}$ in $\S_{\rm int}$ with the
Boltzman factor provided by $\S_{\rm prop}$ one easily obtains the
following one-loop ultraviolet divergent contribution 
to the effective action $\G[\bphi,\bP]$
$$
\G_{\rm div} [\bphi,\bP ; \k_1,\k_2]\,=\, 
\int\! d^2\s\; d\theta\,\Bigl( -{\k_1 \over 4}\, 
\bP^{\a i}D_{\theta} \bP_{\a}^{j}\, {\cal K}^{(1)}_{i j}+
\k_2\, \bP^{\a i}\part_{\a}\bphi^{j}\, {\cal K}^{(2)}_{i j}\Bigr)\; ,
\eqno (11)
$$
with
$$
\eqalignno{
{\cal K}^{(1)}_{i j}\,=&\,
{(\k_1 / \k_2^2) \over 2\pi\epsilon}\Bigl( -{1 \over 2}
 J^m_{\quad\! i} D_k D^k J_{j m} 
- {1\over 3} R _{i j} - {1\over 2} D_k J^m_{\quad\! i} D^k J_{j m}
\Bigr)\; ,\cr         
{\cal K}^{(2)}_{i j}\,=&\,
{(\k_1 / \k_2^2) \over 2\pi\epsilon}\Bigl( -{1 \over 4}
 J^m_{\quad\! i} D_k D^k J_{j m} 
+{7\over 12} R _{i j} - {1\over 8} D_k J^m_{\quad i} D^k J_{j m} 
- {1\over 2} R_{m n} J^n_{\;\; i} J^m_{\quad\! j}\Bigr)\; .& (12)\cr
}
$$
The symbol $\epsilon$ stands for the standard dimensional regularization
regulator. We have taken the manifold $\Sigma$ to be flat. The parameter
$\k_2$ cannot be set to zero, otherwise the free propagator will not
exist.

Eqs. (8), (11) and (12) furnish the tree-level and one-loop 
ultraviolet divergent contributions to the effective action. To 
subtract these divergences we will first  express the bare objects 
$g_{i j}$, $J^i_{\; j}$ and $\bP^{\a i}$ in     
terms of the corresponding renormalized objects 
$$
g_{i j}\,=\, \mu^{-\epsilon} (g^{(r)}_{i j} - \delta g_{i j} )\, ,
J^i_{\; j}\,=\, \mu^{-\epsilon} (J^{(r)\, i}_{\quad\,\; j} - 
\delta J^i_{\; j} )\, ,
\bP^{\a i}\,=\, \bP^{(r)\, \a i} - \delta \bP^{\a i}\; ,
\eqno (13)
$$
and, then, we will substitute eqs. (13) back in eq. (8). 
The symbol $\mu$ of eq. (13) being 
the renormalization scale. As it turns out, renormalization is achieved if
$$
\eqalignno{
&\delta g_{i j}\, =\, {(\k_1 / \k_2^2) \over 2\pi \epsilon }\,\Bigl(
{5\over 6} R_{i j} + {1\over 6} J^k_{\;\, i} R_{kl} J^l_{\; j} 
- {1\over 4} D_kJ^l_{\; i} D^k J_{j l}\Bigr)\; ,\cr
&\delta \bP^{\a i}\, =\,{(\k_1 / \k_2^2) \over 2\pi \epsilon }
\bP^{\a k} \Bigl( {1\over 4} J^m_{\quad\! k} D_l D^l J^i_{\; m}
- {1\over 4} R_k^i -
{1\over 12} J^m_{\quad\! k} R_{m l} J^{l i} + 
{3\over 8} D_m J^l_{\; k} D^m J^i_{\; l}\Bigr)\; .  
& (14)\cr
}
$$
The fact that both $\bP^{\a i}$ and $\bP^{(r)\, \a i}$ ought to be selfdual
leads to the following one-loop constraint
$$
\delta \bP^{\a i}\, =\, \vep^{\a}_{\;\b}
J^{(r)\, i}_{\quad\;\, j}\,\delta \bP^{\b j}+
    \vep^{\a}_{\;\b}\, \delta J^{i}_{\; j}\bP^{(r)\, \b j}
$$
By solving the preceding equation for $\delta J^i_{\; j}$, one obtains
$$
\delta J^i_{\; j}\, =\, {(\k_1 / \k_2^2) \over 2\pi\epsilon}\Bigl(
-{1\over 3} R^i_k J^{(r)\,k}_{\quad\;\,\, j} + 
{1\over 3} J^{(r)\,i}_{\quad\;\, k} 
R^k_j+ {1\over 2} D^k D_k J^{(r)\, i}_{\quad\;\, j}
-{3\over 4} J^i_{\;l}D^m
J^{(r)\, l}_{\quad\;\, k}D_m J^{(r)\, k}_{\quad\;\,\, j}\Bigr)\; .
\eqno (15)
$$
One may show that both the bare tensor $J^i_{\; j}$  and
the renormalized tensor $J^{(r)\, i}_{\quad\;\, j}$ are 
almost complex structures over $M$, modulo two-loop corrections. Indeed,
if, say, $J^{(r)\, i}_{\quad\;\, j}$ is an almost complex structure over M,
the correction $\delta J^i_{\; j}$ in eq. (15) obeys the one-loop consistency
condition
$$ 
J^{(r)\, i}_{\quad\;\, k}\, \delta J^k_j + 
\delta J^i_{\; k}\, J^{(r)\, k}_{\quad\;\; j}\,=\, 0.
$$

  We next introduce the beta function $\b^{g}_{i j}$ of the metric and
the beta function $\b^{J\, i}_{\quad j}$ of the almost 
complex structure as usual:
$$
\b^{g}_{i j}\, =\, \mu\; {\part g^{(r)}_{i j}\over \part \mu},\;
\b^{J\, i}_{\quad j}\, =\,
\mu\; {\part J^{(r)\, i}_{\quad\;\; j}\over \part \mu}  
$$
Eqs. (13), (14) and (15) lead to
$$
\eqalignno{
&\b^{g}_{i j}\, =\,- {(\k_1 / \k_2^2) \over 2\pi}\Bigl(
{5\over 6} R_{i j} +
{1\over 6} J^k_{\;\, i} R_{k l} J^l_{\; j}-{1\over 4}
D_k J^{l}_{\;i} D^k J^{j l}\Bigr)\; ,\cr
&\b^{J\, i}_{\quad j}\, =\,- {(\k_1 / \k_2^2) \over 2\pi}\Bigl(
-{1\over 3} R^i_k J^k_{\;\, j} +
{1\over 3} J^i_{\; k} R^k_j +
{1\over 2} D^k D_k J^i_{\; j}
-{3\over 4} J^i_{\; m}D^k J^m_{\quad\! l}D_k J^l_{\;j}\Bigr)\; .&(16)\cr
}
$$
We have dropped the superscript $r$ from all objects in the preceding 
equations to simplify the notation; they are renormalized objects though.
 
Let us summarize. We have  shown that   both beta functions 
$\b^g_{i j}$ and $\b^{J\, i}_{\quad j}$ depend on the 
gauge parameters $\k_1$ and $\k_2$. These parameters are the 
coefficients of the two $\Q$-exact terms that constitute the classical 
action our model. The beta functions are thus  ``gauge'' dependent 
artifacts. Their value should not affect, therefore, the vacuum 
expectation values of the observables of the model. Notice that
if we set $\k_1\, =\,\k_2\, =\, 1$ in eq. (16) we will retrieve the beta
functions for the action in ref. \cite{\Wittsm}, which were computed in
ref. \cite{\Horne}. If we send $k_1$ to zero, so as to obtain the ``delta
gauge'' action, the beta functions in eq. (16) will also approach zero.
We have thus shown that the beta funstions in ref. \cite{\Horne} can be
connected to the beta functions in ref. \cite{\Birmingham} by 
means of smooth curves.
Also notice that, at variance with topological Yan-Mills 
theories \cite{\Physrep}, the renormalization of the model at hand  
cannot  be carried out by a mere renormalization of its
 ``gauge'' parameters $\k_1$ and $\k_2$. We would also like to 
mention that the counterterm structure we have worked out
is consistent with a Mathai-Quillen interpretation of the renormalized
theory, provided the unregularized model have such an 
interpretation \cite{\Cordes}.

  A final comment. Since we have computed one-loop beta functions,  
we have not paid any attention to a rigorous discussion
of the regularization of the model by means of dimensional 
regularization. Higher loop computations will certainly 
demand such a discussion \cite{\Bos}.

\section{Acknowledgments}
\frenchspacing

C.P. Mart{\'\i}n acknowledges partial finacial support from CICyT.

\section{References}

\frenchspacing

\refno\Wittsm.
E. Witten, Comm. Math. Phys. {\bf 118} 411(1988).

\refno\Physrep.
D. Birmingham, M. Blau, M. Rakowski and G. Thompson, Phys. Rep. 
{\bf 209} 129(1991).

\refno\Horne.
J. H. Horne, Nucl. Phys. {\bf B318} 22(1989).

\refno\Birmingham.
D. Birmingham and M. Rakowski, Modd. Phys. Lett. {\bf A 6} 129(1991).

\refno\Witcoh.
E. Witten, Int. J. Mod. Phys. {\bf A 6} 2775(1991).

\refno\Topmat.
J.M.F. Labastida and P.M. Llatas, Nuc. Phys. {\bf B379} 220(1992).

\refno\Lambert.
N.D. Lambert, Nucl. Phys. {\bf B445} 169(1995).

\refno\Background.
L. Alvarez-Gaum\'e, D.Z. Freedman and S. Mukhi, Ann. Phys. {\bf 134} 85(1981);
S. Mukhi, Nucl. Phys. {\bf B264} 640(1985).

\refno\Cordes.
S. Cordes, G. Moore and S. Ramgoolam, {\it Lectures on 2-d Yang-Mills theory,
equivariant cohomology and topological field theories}, {\tt hep-th/9411210}.

\refno\Bos.
M. Bos, Phys. Lett. {\bf B189} 435(1987); 
H. Osborn, Ann. Phys. {\bf 200} 1(1990).

\bye